\documentclass[11pt,preprint,showpacs,preprintnumbers,amsmath,amssymb]{revtex4}

\usepackage{setspace}
\usepackage{graphicx}
\usepackage{dcolumn}
\usepackage{bm}
\usepackage{multirow}
\usepackage{graphicx}
\usepackage{amssymb}
\usepackage{float}
\usepackage{caption}
\usepackage{rotating}
\usepackage{subfigure}
\usepackage{relsize}
\usepackage{diagbox}
\usepackage{epstopdf}
\usepackage{color}

\usepackage{exscale}
\usepackage{relsize}
\usepackage{dcolumn}
\usepackage{bm}
\usepackage{diagbox}
\usepackage[T1]{fontenc}
\usepackage{caption}

\begin{document}

\title{The properties of the $S$-wave $D_s\bar{D}_s$ bound state}

\author{Jing-Juan Qi 
\footnote{e-mail: qijj@mail.bnu.edu.cn}}
\affiliation{\scriptsize{Junior College, Zhejiang Wanli University, Zhejiang 315101, China}}

\author{Zhen-Yang Wang 
\footnote{Corresponding author, e-mail: wangzhenyang@nbu.edu.cn}}
\affiliation{\scriptsize{Physics Department, Ningbo University, Zhejiang 315211, China}, \scriptsize{Institute of Theoretical Physics, Chinese Academy of Sciences, Beijing 100190, China}}

\author{Zhu-Feng Zhang 
\footnote{e-mail: zhufengzhang@nbu.edu.cn}}
\affiliation{\scriptsize{Physics Department, Ningbo University, Zhejiang 315211, China}}

\author{Xin-Heng Guo 
\footnote{Corresponding author, e-mail: xhguo@bnu.edu.cn}}
\affiliation{\scriptsize{School of Physical Science and Technology, Kunming University, Kunming 650214, China}}

\begin{abstract}
In this work, we investigate possible bound states in the $D_s\bar{D}_s$ system using the Bethe-Salpeter formalism within both the ladder and instantaneous approximations. By numerically solving the Bethe-Salpeter equation with a kernel that incorporates contributions from $\phi$ and $J/\psi$ meson exchanges, we confirm the existence of a loosely bound state. Furthermore, we explore the partial decay widths of the $D_s\bar{D}_s$ bound state into the $D\bar{D}$, $\eta_c\eta$, and $J/\psi\omega$ channels, and observe that these widths are sensitive to the model parameter $\alpha$. Notably, we find the dominant decay channel for the $D_s\bar{D}_s$ bound state to be $D\bar{D}$.
\end{abstract}

\pacs{******}

\maketitle
\section{Introduction}
\label{intro}
Since the discovery of the $X(3872)$ (also known as $\chi_{c1}(3872)$) \cite{Belle:2003nnu}, many charmonium-like states have been reported in experiments. However, the masses of these charmonium-like states do not match the excited states predicted by the relativistic quark model \cite{Godfrey:1985xj}. The structures of these charmonium-like states have thus become the focus of interest and provide a unique window for understanding the nature of the strong force. Interestingly, these states share a common feature in that their masses are close to the thresholds of some hadron pairs, which suggests that they may be loose hadronic molecules \cite{Chen:2022asf,Brambilla:2019esw,Guo:2017jvc,Hosaka:2016pey,Xin:2022bzt,Ji:2022vdj,Abreu:2023rye,Meng:2020cbk,Ji:2022uie,Prelovsek:2020eiw,Xie:2022lyw,Bayar:2022dqa,Peng:2023lfw}.

Not long ago, the LHCb Collaboration reported a new resonant structure, $X(3960)$, with $J^{PC}=0^{++}$, observed in the $D_s^+D_s^-$ mass distribution of the $B^+\rightarrow D_s^+D_s^-K^+$ decay with a significance greater than $12\sigma$ \cite{LHCb:2022vsv}. The mass and width measured by LHCb are
\begin{equation}
    M=3956\pm5\pm10 \,\text{MeV},\quad\quad\Gamma=43\pm13\pm8\, \text{MeV}.
\end{equation}
This structure lies just above the $D_s^+D_s^-$ threshold and represents an excellent candidate for a $D_s^+D_s^-$ hadronic molecule. The LHCb Collaboration also reported a scalar state, $X_0(3930)$, in the $D^+D^-$ mass distribution of the $B^+\rightarrow D^+D^-K^+$ decay \cite{LHCb:2020bls,LHCb:2020pxc} in 2020, with a mass and width of
\begin{equation}
    M=3924\pm2 \,\text{MeV},\quad\quad\Gamma=17\pm5\, \text{MeV},
\end{equation}
which lies 10 MeV below the $D_s^+D_s^-$ threshold. The Particle Data Group classifies these as the same particle as the structure first observed by the Belle experiment in the $\omega J/\psi$ invariant mass spectrum of the $B \to K \omega J/\psi$ process \cite{Belle:2004lle}, referred to as X(3915) (or $\chi_{c0}(3915)$) \cite{ParticleDataGroup:2024cfk}.

There have been extensive theoretical studies of $X(3915)$, with several investigating its possible assignment as a charmonium state. Within an unquenched quark model, Ref. \cite{Duan:2020tsx} finds the mass and decay width of $\chi_{c0}(2P)$ to be consistent with measurements of $X(3915)$. Using a rescattering mechanism to compute $\mathcal{B}[B \rightarrow K\chi_{c0}(2P)]$, Ref. \cite{Duan:2021bna} argues that the $\chi_{c0}(2P)$ state plays a crucial role in the LHCb measurement of $B \rightarrow K D\bar{D}$. Coupled-channel analyses indicate that assigning $X(3960)$ to the $\chi_{c0}(2P)$ charmonium state cannot be excluded \cite{Man:2024mvl}. Because $\chi_{c0}(2P)$ predominantly decays to $D\bar{D}$, Ref. \cite{Qian:2023taw} proposes $e^+e^- \rightarrow \omega D\bar{D}$ as an ideal process to identify this state. In contrast, quark-model calculations employing the $^3P_0$ decay model find that, although the mass of $\chi_{c0}(2^3P_0)$ is compatible with $X(3915)$, its predicted strong decay width is much larger than the experimental value \cite{Yang:2009fj}. Guo and Meissner \cite{Guo:2012tv}, analyzing BaBar data, likewise conclude that $X(3915)$ is difficult to interpret as $\chi_{c0}(2P)$, a view supported by Ref. \cite{Olsen:2014maa}. Within a constituent quark framework, Ref. \cite{Ortega:2017qmg} finds $X(3915)$ to be dominantly molecular, with a bare $q\bar{q}$ probability below $45\%$, and favors a $J^{PC}=2^{++}$ assignment. Finally, an analysis of the $D\bar{D}$ invariant-mass distribution in $e^+e^- \rightarrow J/\psi D\bar{D}$ measured by the Belle Collaboration reveals no clear enhancement above the $D\bar{D}$ threshold to support the existence of a $\chi_{c0}(2P)$ state \cite{Wang:2019evy}.

Interpreting X(3915) as a $D_s\bar{D}_s$ molecular state is another popular approach \cite{Ji:2022vdj,Abreu:2023rye,Meng:2020cbk,Ortega:2017qmg,Ji:2022uie,Prelovsek:2020eiw,Xie:2022lyw,Bayar:2022dqa,Li:2015iga,Chen:2016ncs,Ding:2021igr,Mutuk:2022ckn,Chen:2023eix,Dong:2021juy,Peng:2023lfw,Liu:2008mi,Zhang:2009st,Liu:2009qhy,Liu:2017mrh}. By studying the interaction of the $D\bar{D}$ and $D_s\bar{D}_s$ coupled channels in $B^-\rightarrow K^-J/\psi\omega$ decay \cite{Abreu:2023rye} and $B^+\rightarrow D^+D^-(D_s^+D_s^-)K^+$ decay \cite{Bayar:2022dqa}, a $D_s\bar{D}_s$ bound state was found around 3930 MeV, which couples strongly to $D_s\bar{D}_s$, with no extra resonance signal at 3960 MeV. Such a state was also predicted in lattice QCD simulations \cite{Prelovsek:2020eiw}, with a shallow binding energy of about 6.2 MeV. The authors of Ref. \cite{Ji:2022uie} employed an effective field theory based on heavy quark spin symmetry and found that the $D_s^+D_s^-$ invariant mass distribution data can be well described by either a bound or a virtual state below the $D_s^+D_s^-$ threshold, with a pole mass of (3928$\pm$3) MeV. The existence of a $D_s\bar{D}_s$ bound state from $B$ decay processes was also supported by effective Lagrangian approaches \cite{Xie:2022lyw}. The $D_s\bar{D}_s$ system can also form a bound state in a contact-range theory \cite{Peng:2023lfw}. Based on the vector-meson-dominance model \cite{Dong:2021juy}, the potential from $\phi$ exchange is insufficient to form a $D_s\bar{D}_s$ bound state; instead, a shallow virtual state was obtained with a virtual energy roughly in the range [4.7, 35.5] MeV. No $D_s\bar{D}_s$ bound state was found in a potential model including $\phi$ exchange contributions \cite{Liu:2017mrh}. In Ref. \cite{Ding:2021igr}, the effects of both $\phi$ and $J/\psi$ exchange were considered, and it was found that the $D_s\bar{D}_s$ system could exist as a bound state within the quasipotential Bethe-Salpeter equation approach. Therefore, elucidating the properties of the $D_s\bar{D}_s$ bound state is important for understanding the nature of $X(3930)$ and $X(3960)$.

The purpose of the present work is to apply the Bethe-Salpeter (BS) equation to investigate the possibility of a $D_s\bar{D}_s$ bound state when contributions from $\phi$ and $J/\psi$ meson exchanges are considered. We will further investigate the partial decay widths of a possible $D_s\bar{D}_s$ bound state.

This work is organized as follows. In Sec. \ref{BS}, we review the basic BS formalism for a system of two pseudoscalar particles. The formalism for the partial decay widths of the $D_s\bar{D}_s$ bound state to $D\bar{D}$, $\eta_c\eta$, and $J/\psi\omega$ final states will be presented in Sec. \ref{Deacy formalism}. Numerical results are presented in Sec. \ref{Num}. A brief summary is given in the final section.

\section{Bethe-Salpeter equation for the $D_s\bar{D}_s$ system}
\label{BS}

In this section, we derive the BS equation for the $S$-wave $D_s\bar{D}_s$ system. The BS wave function for the $S$-wave $D_s\bar{D}_s$ bound state with total momentum $P$ is defined as
\begin{equation}
    \chi_P(x_1,x_2)=\langle0|TD_s(x_1)\bar{D}_s(x_2)|P\rangle=e^{-iPX}\int\frac{d^4p}{(2\pi)^4}\chi_P(p)e^{-ipx},
\end{equation}
where the coordinate $X$ denotes the center of mass, while $p$ and $x$ represent the relative momentum and relative coordinate of the $D_s$ and $\bar{D}_s$ pair, respectively. We define $\lambda_i = m_{D_s(\bar{D}_s)}/(m_{D_s} + m_{\bar{D}_s}) = 1/2$ (for $i = 1,2$). The individual momenta of $D_s$ and $\bar{D}_s$ can then be expressed in terms of $P$ and $p$ as $p_1 = \lambda_1 P + p$ and $p_2 = \lambda_2 P - p$, respectively.

The $S$-wave $D_s\bar{D}_s$ bound state wave function $\chi_P(p)$ satisfies the following BS equation:
\begin{equation}\label{BS eq.}
    \chi_P(p)=S_{D_s}(p_1)\int\frac{d^4q}{(2\pi)^4}K_P(P,p,q)\chi_P(q)S_{\bar{D}_s}(p_2),
\end{equation}
where $S_{D_s}(p_1)$ and $S_{\bar{D}_s}(p_2)$ are the propagators for $D_s$ and $\bar{D}_s$, respectively, and $K_P(P,p,q)$ is the kernel which can be derived from the four-point Green function.

The interaction kernel of the $D_s\bar{D}_s$ system can be obtained through the exchange of $\phi$ and $J/\psi$ mesons. Since the contribution from $J/\psi$ exchange was found to be significant in the $D\bar{D}^\ast$ interaction for the production of the $Z_c(3900)$ \cite{Aceti:2014uea,He:2013nwa}, and its effect is likewise crucial for the $D\bar{D}$ system to form a bound state \cite{Li:2022shq}, we include the contribution of $J/\psi$ exchange in the current work. The Lagrangian can be constructed based on chiral symmetry and heavy quark symmetry:
\begin{equation}
\begin{split}
  \mathcal{L}_{D_sD_s\phi}&=-ig_{D_sD_s\phi}(D_s^\dag \partial^\mu D_s-\partial^\mu D_s^\dag D_s)\phi_\mu,\\
  \mathcal{L}_{D_sD_sJ/\psi}&= -ig_{D_sD_s J/\psi}(D_s^\dag \partial^\mu D_s-\partial^\mu D_s^\dag D_s)J/\psi_\mu.
  \end{split}
\end{equation}
The magnitudes of the coupling parameters are very important for the possibility of forming a bound state. For the coupling constant \(g_{D_sD_s\phi}\), the results from the SU(3) relation (\(g_{D_sD_s\phi}=\frac{m_{D_s}}{m_D}g_{DD\rho}\)) \cite{Cheng:2004ru} and from QCDSR \cite{Bracco:2012mp,Wang:2007zm,OsorioRodrigues:2017aqv} vary widely, as listed in Table \ref{constant}. The coupling constant \(g_{D_sD_sJ/\psi}\) has been obtained through flavor SU(3) symmetry (\(g_{D_sD_sJ/\psi} = g_{DDJ/\psi}\)) \cite{Liu:2006dq}, and from the relation with the gauge coupling \(g_2\) (\(g_{D_sD_sJ/\psi}=2g_2m_D\sqrt{m_{J/\psi}}\) with \(g_2=\sqrt{m_{J/\psi}}/(2m_Df_{J/\psi})\) and \(f_{J/\psi}=405\) MeV) \cite{Ding:2021igr}.

\begin{table}[h]
\renewcommand{\arraystretch}{1.2}
\centering
\caption{The value of the coupling constant $g_{D_sD_s\phi}$.}\label{constant}
\begin{tabular*}{\textwidth}{@{\extracolsep{\fill}}lcccc}
\hline
\hline
     & \cite{Cheng:2004ru} &\cite{Bracco:2012mp}  & \cite{Wang:2007zm}  &  \cite{OsorioRodrigues:2017aqv}   \\ 
\hline
$g_{D_sD_s\phi} $ & 3.89 & $1.90^{+0.17}_{-0.13}$ & $1.45 \pm 0.34$   &$2.08^{+0.41}_{-0.43}$                    \\
\hline
\hline
\end{tabular*}
\end{table}

At the tree level, the $t$-channel interaction kernel for the BS equation in the so-called ladder approximation is
\begin{equation}\label{kernel}
K(P,p,q)=(2\pi)^4\delta^4(p_1+p_2-q_1-q_2)g_{D_sD_s\phi}^2(p_1^\mu+q_1^\mu)(p_2^\nu+q_2^\nu)\Delta_{\mu\nu}(p_1-q_1),
\end{equation}
where $\Delta_{\mu\nu}(p_1-q_1)$ represents the propagator for the $\phi$ or $J/\psi$ meson.

Since the strong interaction vertices are determined by the physical particles and the off-shell exchanged particles, it is necessary to introduce a form factor $F(k)$ to account for the off-shell effects of the $t$-channel exchanged particles. The form factor is defined as
\begin{equation}\label{f.f.}
    F(k,m_V)= \frac{\Lambda^2-m_V^2}{\Lambda^2-k^2},
\end{equation}
where $k$ is the momentum of the exchanged particle, $m_V$ is its physical mass, and $\Lambda$ is the cutoff in the form factor. The value of the cutoff $\Lambda$ should be close to the physical mass of the exchanged particle. It can be reparameterized as $\Lambda=m_V+\alpha \Lambda_{\text{QCD}}$, where $\Lambda_{\text{QCD}}$ is the QCD scale (approximately 220 MeV) and $\alpha$ is expected to be of order unity. The value of $\alpha$ depends on the exchanged and external particles involved in the strong interaction vertex and cannot be obtained from first principles.

To solve the BS Eq. (\ref{BS eq.}), we use the instantaneous approximation in the kernel. In this approximation, the energy exchanged between the constituent particles of the bound system is neglected. Studies from lattice QCD \cite{Prelovsek:2020eiw}, effective field theory \cite{Ji:2022uie}, and contact-range theory \cite{Peng:2023lfw} suggest that the $D_s\bar{D}_s$ system has a very small binding energy. This indicates that the binding of the constituent particles is weak, making it reasonable to employ the instantaneous approximation in the kernel of the BS equation.

Substituting Eqs. (\ref{kernel}) and (\ref{f.f.}) into Eq. (\ref{BS eq.}) and applying the instantaneous approximation, in the center-of-mass frame of the bound state ($P=(M,\mathbf{0})$), we obtain:
\begin{equation}
\begin{split}\label{four-BS}
\chi_P(p)
&=\frac{i}{\left[(\lambda_1M+p_0)^2-\omega_1^2+i\epsilon\right]\left[(\lambda_2M-p_0)^2-\omega_2^2+i\epsilon\right]}\\
&\int\frac{d^4q}{(2\pi)^4}g_{D_sD_sV}^2\frac{4\lambda_1\lambda_2M^2+(\mathbf{p}+\mathbf{q})^2+\frac{(\mathbf{p}^2-\mathbf{q}^2)^2}{m_V^2}}{-(\mathbf{p}-\mathbf{q})^2-m_V^2}F(\mathbf{k})^2\chi_P(q),\\
\end{split}
\end{equation}
for each exchanged meson with mass \(m_V\) and coupling \(g_{D_sD_sV}\), \(\omega_{1(2)}\) is the energy defined as \(\sqrt{m^2_{D_s(\bar{D}_s)}+\mathbf{p}^2}\).

Then, performing the integration over \(p_0\) on both sides (applying the residue theorem on the right-hand side) and over \(q_0\) on the right-hand side, Eq. (\ref{four-BS}) becomes
\begin{equation}\label{three-BS}
\begin{split}
\chi_P(\mathbf{p})
&= \frac{E_1+E_2}{2E_1E_2(M-E_1-E_2)(M+E_1+E_2)}\\
&\int\frac{d^3\mathbf{q}}{(2\pi)^3}g_{D_sD_sV}^2\frac{4\lambda_1\lambda_2M^2+(\mathbf{p}+\mathbf{q})^2+\frac{(\mathbf{p}^2-\mathbf{q}^2)^2}{m_V^2}}{-(\mathbf{p}-\mathbf{q})^2-m_V^2}F(\mathbf{k})^2\chi_P(\mathbf{q}).\\
\end{split}
\end{equation}
To solve the integral equation (\ref{three-BS}), we need to complete the azimuthal integration and discretize the integration region into $n$ pieces (with $n$ sufficiently large). This discretization transforms the integral equation into an eigenvalue equation for the $n$-dimensional vector $\chi_P(|\mathbf{p}|)$. Then we can solve the BS equation numerically.

\section{The partial decay widths of the $D_s\bar{D}_s$ bound state}
\label{Deacy formalism}
In this section, we apply the obtained BS wave functions to calculate the partial decay widths of the $D_s\bar{D}_s$ bound state. Prior to this, it is necessary to obtain the normalized BS wave function, which corresponds to determining the physical wave function. The normalization condition for the BS wave function is given by
\begin{equation}\label{Nor BS}
    i\int\frac{d^4pd^4q}{(2\pi)^4}\bar{\chi}_P(p)\frac{\partial}{\partial P^0}\left[I_P(p,q)+\bar{K}_P(p,q)\right]\chi_P(q)=2E_P, \, P^0=E_P,
\end{equation}
where $I_P(p,q)=(2\pi)^4\delta(p-q)S_{D_s}^{-1}(p_1)S_{\bar{D}_s}^{-1}(p_2)$. The energy of the bound state is denoted as $E_P$, and in the rest frame of the bound state, we have $P^0=E_P=M$.

After carrying out the $p_0$-integrations using appropriate contours and completing the azimuthal integration for the normalization equation (\ref{Nor BS}), we obtain
\begin{equation}
\begin{split}
&-\int\frac{d|\mathbf{p}|}{(2\pi)^4}\frac{8\omega_1\omega_2P_0|\mathbf{p}|^2}{\omega_1+\omega_2}\bar{\chi}_P^2(|\mathbf{p}|)+\frac{4\lambda_1\lambda_2P_0}{(2\pi)^6}\int d|\mathbf{p}|d|\mathbf{q}||\mathbf{p}||\mathbf{q}|g^2_{D_sD_sV}\bar{\chi}_P(|\mathbf{p}|)T\chi_P(|\mathbf{p}|)=2E_P.
\end{split}
\end{equation}
with
\begin{equation}
T=\frac{\Lambda^2-m_V^2}{\Lambda^2+(|\mathbf{p}|+|\mathbf{q}|)^2}-
\frac{\Lambda^2-m_V^2}{\Lambda^2+(|\mathbf{p}|-|\mathbf{q}|)^2}+\ln\frac{\Lambda^2+(|\mathbf{p}|-|\mathbf{q}|)^2}{\Lambda^2+(|\mathbf{p}|+|\mathbf{q}|)^2}-\ln\frac{m_V^2+(|\mathbf{p}|-|\mathbf{q}|)^2}{m_V^2+(|\mathbf{p}|+|\mathbf{q}|)^2}.
\nonumber
\end{equation}
Then, we can obtain the normalized BS wave function. If the $D_s\bar{D}_s$ bound state wave function obtained in the previous section does not satisfy this normalization equation (the left side of the normalization equation gives some constant $\mathcal{C}^2\neq 1$), one must replace $\chi_P(|\mathbf{p}|)\rightarrow\chi_P(|\mathbf{p}|)/|\mathcal{C}|$ to ensure the correct normalization of the BS wave function.

Now, we proceed to study the decay widths of the $D_s\bar{D}_s$ bound states. The relevant interaction Lagrangians for the $D_s\bar{D}_s$ bound state decaying to $D\bar{D}$, $\eta_c\eta$, and $J/\psi\omega$ are \cite{Chen:2016ncs}
\begin{equation}\label{lagran}
    \begin{split}
\mathcal{L}_{\mathcal{D}\mathcal{D}\mathcal{V}}&=-ig_{\mathcal{D}\mathcal{D}\mathcal{V}}(\mathcal{D}\partial_\mu \mathcal{D}^\dag-\partial_\mu \mathcal{D} \mathcal{D}^\dag)\mathcal{V}^\mu+c.c.,\\
\mathcal{L}_{\eta_c\mathcal{D}^\ast \mathcal{D}}&=-ig_{\mathcal{D}\eta_c\mathcal{D}^\ast}(\partial_\mu\eta_c\mathcal{D}-\eta_c\partial_\mu \mathcal{D})\mathcal{D}^{\ast\mu\dag}+c.c.,\\
\mathcal{L}_{\eta \mathcal{D}^\ast \mathcal{D}}&=-ig_{\eta \mathcal{D}^\ast \mathcal{D}}(\mathcal{D}\partial^\mu\eta \mathcal{D}_{\mu}^{\ast\dag}-\mathcal{D}_{\mu}^{\ast}\partial^\mu\eta \mathcal{D}^\dag),\\
\mathcal{L}_{\mathcal{D}\mathcal{D}J/\psi}&=ig_{\mathcal{D} \mathcal{D}J/\psi}(\partial^\mu \mathcal{D} \mathcal{D}^\dag-\mathcal{D}\partial^\mu \mathcal{D}^\dag)J/\psi_\mu,\\
\mathcal{L}_{\mathcal{D} \mathcal{D}\omega}&=ig_{\mathcal{D} \mathcal{D}\mathcal{V}}(\partial_\mu \mathcal{D} \mathcal{D}^\dag-\mathcal{D}\partial_\mu \mathcal{D}^\dag)\mathcal{V}^\mu,\\
\mathcal{L}_{\mathcal{D}^\ast \mathcal{D} J/\psi}&=g_{J/\psi \mathcal{D}^\ast \mathcal{D}}\epsilon^{\mu\nu\alpha\beta}\partial_\mu J/\Psi_\nu\left(\mathcal{D}_\alpha^\ast\overleftrightarrow{\partial}_\beta \mathcal{D}^\dag-\mathcal{D}\overleftrightarrow{\partial}_\beta \mathcal{D}^{\ast\dag}_\alpha\right),\\
\mathcal{L}_{\mathcal{D}^\ast_s \mathcal{D}V}&=2f_{\mathcal{D}^\ast \mathcal{D}V}\epsilon^{\mu\nu\alpha\beta}\partial_\mu \mathcal{V}_\nu\left(\mathcal{D}_\alpha^\ast\overleftrightarrow{\partial}_\beta \mathcal{D}^\dag-\mathcal{D}\overleftrightarrow{\partial}_\beta \mathcal{D}^{\ast\dag}_\alpha\right),\\
    \end{split}
\end{equation}
where $\mathcal{D}=\{\mathcal{D}^{(\ast)0},\mathcal{D}^{(\ast)+},\mathcal{D}_s^{(\ast)+}\}$, and $\mathcal{V}$ and $\mathcal{P}$ are the matrices of vector and pseudoscalar mesons, respectively, which have the following forms:
\begin{eqnarray}\label{vmatrix}
\mathcal{V}&=&\left(\begin{array}{ccc}
\frac{\rho^{0}}{\sqrt{2}}+\kappa\omega+\zeta\phi&\rho^{+}&K^{\ast+}\\
\rho^{-}&-\frac{\rho^{0}}{\sqrt{2}}+\kappa\omega+\zeta\phi&
K^{\ast0}\\
K^{\ast-} &\bar{K}^{\ast0}&\lambda\omega+\sigma\phi
\end{array}\right),
\end{eqnarray}
\begin{eqnarray}\label{pmatrix}
\mathcal{P}&=&\left(\begin{array}{ccc}
\frac{\pi^{0}}{\sqrt{2}}+\tau\eta+\xi\eta'&\pi^{+}&K^{+}\\
\pi^{-}&-\frac{\pi^{0}}{\sqrt{2}}+\tau\eta+\xi\eta'&
K^{0}\\
K^{-} &\bar{K}^{0}&\gamma\eta+\delta\eta'
\end{array}\right).
\end{eqnarray}
In Eq. (\ref{vmatrix}), the physical $\omega$ and $\phi$ are mixtures of the ideal isospin eigenstates $\omega^I$ and $\phi^I$ with the mixing angle $\theta_v$, where $\kappa = \cos\theta_v / \sqrt{2}$, $\zeta = \sin\theta_v / \sqrt{2}$, $\lambda = -\sin\theta_v$, and $\sigma = \cos\theta_v$. Similarly, in Eq. (\ref{pmatrix}), $\eta$ and $\eta'$ are mixtures of the ideal isospin eigenstates $\eta^I$ and $\eta^{'I}$ with the mixing angle $\theta_p$, where $\tau = (\cos\theta_p - \sqrt{2}\sin\theta_p) / \sqrt{6}$, $\xi = (\sin\theta_p + \sqrt{2}\cos\theta_p) / \sqrt{6}$, $\gamma = (-2\cos\theta_p - \sqrt{2}\sin\theta_p) / \sqrt{6}$, and $\delta = (-2\sin\theta_p + \sqrt{2}\cos\theta_p) / \sqrt{6}$. In this work, we adopt the central value $\theta_v = (3.4 \pm 0.2)^\circ$, as determined in Refs. \cite{Benayoun:1999fv,Kucukarslan:2006wk,Dolinsky:1991vq}. We also take $\theta_p = -19.1^\circ$ as in Refs. \cite{MARK-III:1988crp,DM2:1988bfq}. The relevant coupling constants can be estimated using the heavy quark limit and chiral symmetry.
\begin{equation}
    \begin{split}
        &g_{J/\psi D_sD_s}=2g_2\sqrt{m_{J/\psi}}m_{D_s},\,\, g_{J/\psi D_s^\ast D_s}=2g_2\sqrt{m_{D_s}^\ast m_{D_s}/m_{J/\psi}},\,\,
        g_{\eta_cD_s^\ast D_s}=2g_2\sqrt{m_{\eta_c}m_{D_s}^\ast m_{D_s}},\\
        &g_{D_s^\ast D_s\eta}=2g\gamma\sqrt{m_{D_s^\ast}m_{D_s}}/f_\pi,\,\,g_{D_sD_s\omega}=\lambda\beta_Vg_V/\sqrt{2},\,\,g_{D_s^\ast D_s\omega}=\lambda\lambda_Vg_V/\sqrt{2},\,\,g_{D_sDK^\ast}=\beta_Vg_V/\sqrt{2},
    \end{split}
\end{equation}
The relevant coupling constants in Eq.~(15) of our manuscript can be estimated in the heavy quark limit and with chiral symmetry, where the coupling constants of the $S$-wave charmonia and charmed mesons can be determined by
\begin{equation}
\begin{split}\nonumber
    g_{J/\psi D_sD_s}&=2g_2\sqrt{m_{J/\psi}}m_{D_s},\,\,g_{J/\psi D_s^\ast D_s}=2g_2\sqrt{m_{D_s^\ast} m_{D_s}/m_{J/\psi}},\\ g_{\eta_c D_s^\ast D_s}&=2g_2\sqrt{m_{\eta_c}m_{D_s^\ast} m_{D_s}},
\end{split}
\end{equation}
where the gauge coupling $g_2$ relates to the decay constant of $J/\psi$ by $g_2=\sqrt{m_{J/\psi}}/(2m_Df_{J/\psi})$, which is estimated from the leptonic decay width of $J/\psi$ \cite{Chen:2016ncs}. The other coupling constants can be estimated by
\begin{equation}
    \begin{split}\nonumber
        g_{D_s^\ast D_s\eta}&=2g\gamma\sqrt{m_{D_s^\ast}m_{D_s}}/f_\pi,\,\,g_{D_sD_s\omega}=\lambda\beta_Vg_V/\sqrt{2},\\
        g_{D_s^\ast D_s\omega}&=\lambda\lambda_Vg_V/\sqrt{2},\,\,g_{D_sDK^\ast}=\beta_Vg_V/\sqrt{2},
    \end{split}
\end{equation}
where $\beta_V$ can be fixed to 0.9 by vector meson dominance, $\lambda_V$ is obtained from light-cone sum rules and lattice QCD as $0.56\, \rm{GeV}^{-1}$. Imposing the Kawarabayashi-Suzuki-Riazuddin-Fayyazuddin relations, $g_V = m_\rho/f_\pi$, with $f_\pi = 0.132$ GeV (the pion decay constant), and $g = 0.59$ (estimated from the partial width for $D^\ast\rightarrow D\pi$) \cite{Isola:2003fh}. Since we have accounted for variations in the coupling constant $g_{D_sD_s\phi}$ from different approaches while solving the BS equation—and these variations span a wide range—we proceed with the analysis.

With the effective Lagrangians in Eq. (\ref{lagran}), the lowest-order amplitudes for the decays of the $D_s\bar{D}_s$ bound state to $D\bar{D}$, $\eta_c\eta$, and $J/\psi\omega$ are
\begin{align}\label{DDAM}
    \mathcal{M}_{D\bar{D}}=&g^2_{D_sDK^\ast}\int\frac{d^4p}{(2\pi)^4}(p_1+p'_1)_\mu(p_2+p'_2)_\nu\Delta_{K^\ast}^{\mu\nu}(k)F(k)^2\chi_P(p'),\\
    \mathcal{M}_{\eta_c\eta}=&g_{\eta_cD_s^\ast D_s}g_{D_s^\ast D_s\eta}\int\frac{d^4q}{(2\pi)^4}\big[(p_1+p'_1)_\mu p'_{2\nu}\Delta_{D_s^\ast}^{\mu\nu}(k)F(k)^2|_{k=p'_1-p_1}\nonumber\\
    &+(p_2+p'_1)_\mu p'_{2\nu}\Delta_{D_s^\ast}^{\mu\nu}(k)|_{k=p_1-p'_2}F(k)^2\big]\chi_P(p'),\\
\mathcal{M}_{J/\psi\omega}=& \mathcal{M}_{J/\psi\omega}^{D_s}+\mathcal{M}_{J/\psi\omega}^{D_s^\ast}\nonumber\\
=&g_{J/\psi D_s D_s}g_{D_s D_s\omega}\varepsilon^{J/\psi}_\mu(p'_1)\varepsilon^\omega_\nu(p'_2)\int\frac{d^4p}{(2\pi)^4}\big[(p_1-k)^\mu (p_2+k)^\nu\Delta_{D_s}(k)F(k)^2|_{k=p'_1-p_1}\nonumber\\
&+(p_1+k)^\mu (p_2-k)^\nu\Delta_{D_s}(k)F(k)^2\chi_P(p')|_{k=p_1-p'_2}\nonumber\\
&+2g_{J/\psi D_s^\ast D_s}f_{D_s^\ast D_s\omega}\epsilon_{\mu\nu\alpha\beta}\epsilon_{\kappa\lambda\rho\tau}\int\frac{d^4p}{(2\pi)^4}\big[\epsilon_{J/\psi}^\nu(p'_1)\epsilon_\omega^\lambda(p'_2)p_1^{'\mu}(p_1-k)_\beta p_2^{'\kappa}(p_2+k)^\tau\Delta_{D_s^\ast}^{\alpha\rho}(k)F(k)^2|_{k=p'_1-p_1}\nonumber\\
&+\epsilon_{J/\psi}^\lambda(p'_1)\epsilon_{\omega}^\nu(p'_2)p_2^{'\mu}(p_1+k)_\beta p_1^{'\kappa}(p_2-k)^\tau\Delta_{D_s^\ast}^{\alpha\rho}(k)F(k)^2|_{k=p_1-p'_2}\big]\chi_P(p'),
\end{align}
respectively, where $p'_i (i=1,2)$ is the momentum of the $i$-th particle in the final state. In the rest frame of the bound state, the momenta of the two particles in the final state can be taken as $p'_1=(E_1,\mathbf{p}')$, $p'_2=(E_2,-\mathbf{p}')$ with $E_1=\sqrt{\mathbf{p}^{'2}+m_1^2}$ and $E_2=\sqrt{\mathbf{p}^{'2}+m_2^2}$. We define $p'\equiv \lambda_2 p'_1-\lambda_1p'_2$, which is not the relative momentum of the final particles; then, $p'=(\lambda_2E_1-\lambda_1E_2,\mathbf{p}')$.

In the rest frame, the two-body decay width of the bound state can be written as
\begin{equation}
    d\Gamma=\frac{1}{32\pi^2}|\mathcal{M}|^2\frac{|\mathbf{p}'|}{M^2}d\Omega,
\end{equation}
where
\begin{equation}
    |\mathbf{p}'|=\frac{1}{2M}\sqrt{\lambda(M^2,m_1^2,m_2^2)},
\end{equation}
the Källén function is given by $\lambda(a,b,c)=a^2+b^2+c^2-2ab-2ac-2bc$.

\section{Numerical results}
\label{Num}

The coupling constants and form factors associated with strong interaction vertices play a crucial role in studying possible bound states of the $D_s\bar{D}_s$ system and in estimating partial decay widths of the $D_s\bar{D}_s$ bound state into $D\bar{D}$, $\eta_c\eta$, and $J/\psi\omega$ final states. As shown in Table \ref{constant}, the magnitude of $g_{D_sD_s\phi}$ varies considerably depending on whether it is obtained from the SU(3) relation or from QCD sum rules. In our work, we consider the range of $g_{D_sD_s\phi}$ values from the minimum to the maximum presented in Table \ref{constant}.

Since the exchanged particle is not on-shell, it is necessary to introduce a form factor $F(k)$ to account for the off-shell effect of the $t$-channel exchanged particle. The cutoff $\Lambda$ in the form factor $F(k)$ should be close to the physical mass of the exchanged particle. We take $\Lambda=m+\alpha\Lambda_{\text{QCD}}$, where $\alpha$ depends on the exchanged particle and the external particles involved in the strong interaction vertex. The parameter $\alpha$ is expected to be of order unity and cannot be determined from first principles. In this work, we identify all possible values of $\alpha$ for which the $D_s\bar{D}_s$ system can form a bound state.

\begin{figure}[ht]
\centering
    \rotatebox{0}{\includegraphics*[width=0.4\textwidth]{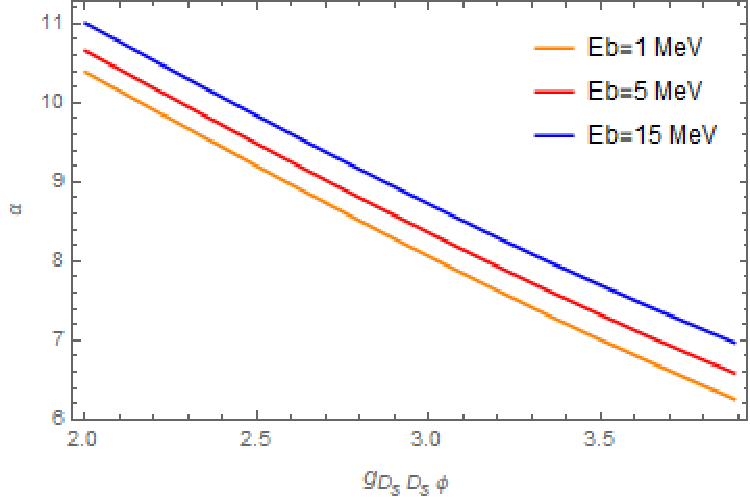}}
    \caption{The possible bound states in the $D_s\bar{D}_s$ system.}
  \label{DsDs}
\end{figure}

We then solve the BS equation (\ref{three-BS}) numerically by discretizing it into a matrix eigenvalue equation using the Gaussian quadrature method. For each trial set of values of the coupling constant $g_{D_sD_s\phi}$, the parameter $\alpha$, and the binding energy $E_b$ of the $D_s\bar{D}_s$ system (defined as $E_b=m_{D_s}+m_{\bar{D}_s}-E$ and varying from 0 to 15 MeV), we obtain all eigenvalues of this equation. The set of values of $g_{D_sD_s\phi}$, $\alpha$, and $E_b$ that yields the eigenvalue closest to 1.0 is selected. We find that for a very small coupling constant $g_{D_sD_s\phi}$, a very large parameter $\alpha$ is required for a bound state to exist. Therefore, we choose $g_{D_sD_s\phi}$ in the interval of (2–3.89) for our analysis. The values of $g_{D_sD_s\phi}$ and $\alpha$ for possible bound states with binding energies $E_b=1$ MeV, $E_b=5$ MeV, and $E_b=15$ MeV are presented in Fig. \ref{DsDs}. From Fig. \ref{DsDs}, we observe that for a given coupling constant $g_{D_sD_s\phi}$, a larger binding energy requires a larger value of the parameter $\alpha$. Additionally, our study reveals that the $D_s\bar{D}_s$ system cannot form a bound state if only the contribution of $\phi$ or $J/\psi$ exchange is considered.

\begin{figure}[htbp]
\centering
\subfigure[]{
\begin{minipage}[t]{0.3\linewidth}
\centering
\includegraphics[width=2.0in]{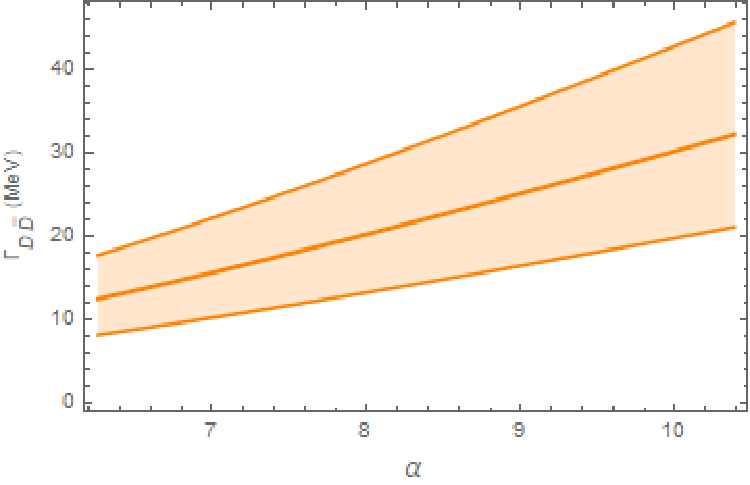}
\end{minipage}
}
\subfigure[]{
\begin{minipage}[t]{0.3\linewidth}
\centering
\includegraphics[width=2.0in]{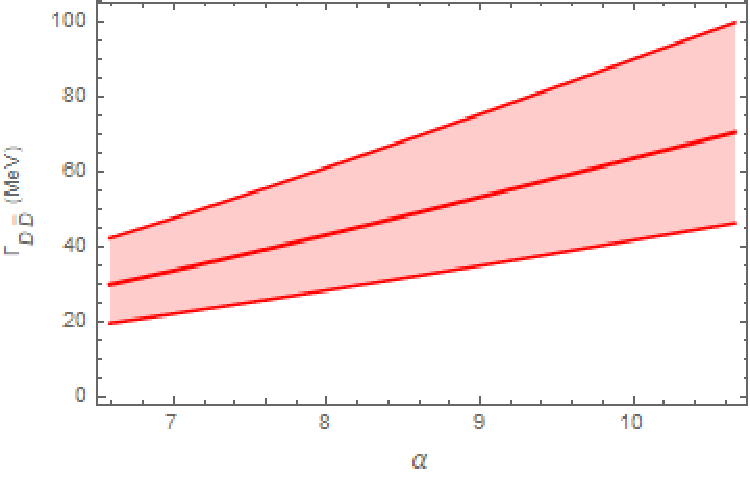}
\end{minipage}
}
\subfigure[]{
\begin{minipage}[t]{0.3\linewidth}
\centering
\includegraphics[width=2.0in]{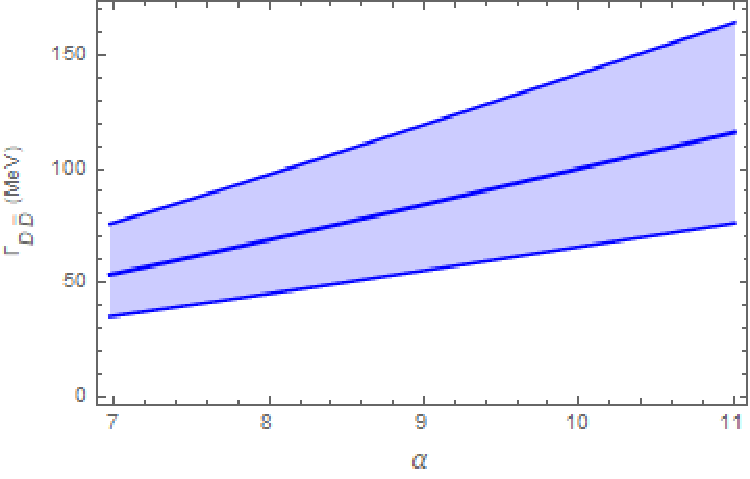}
\end{minipage}
}
\centering
\caption{The particle decay width of the $D_s\bar{D}_s$ bound state to $D\bar{D}$ is shown for $E_b$ = 1 MeV (a), $E_b$ = 5 MeV (b), and $E_b$ = 15 MeV (c) across the allowed parameter $\alpha$.}
\label{DD}
\end{figure}

\begin{figure}[htbp]
\centering
\subfigure[]{
\begin{minipage}[t]{0.3\linewidth}
\centering
\includegraphics[width=2.0in]{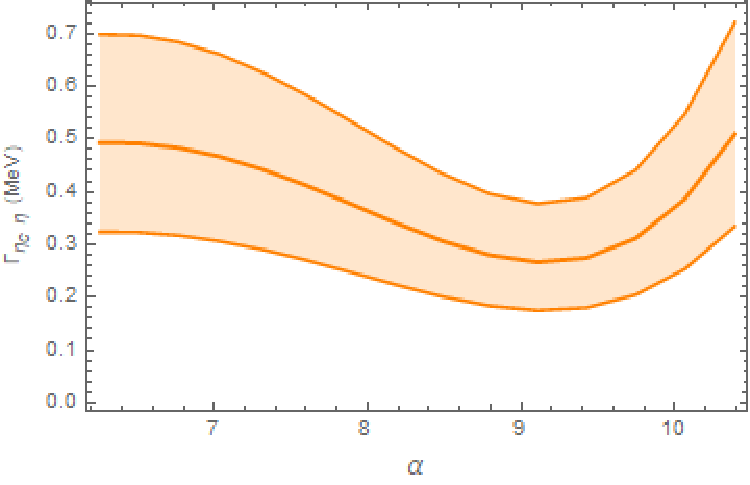}
\end{minipage}
}
\subfigure[]{
\begin{minipage}[t]{0.3\linewidth}
\centering
\includegraphics[width=2.0in]{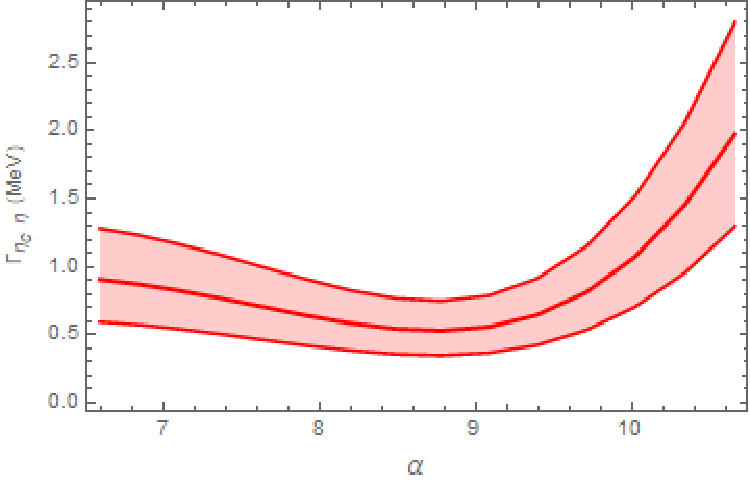}
\end{minipage}
}
\subfigure[]{
\begin{minipage}[t]{0.3\linewidth}
\centering
\includegraphics[width=2.0in]{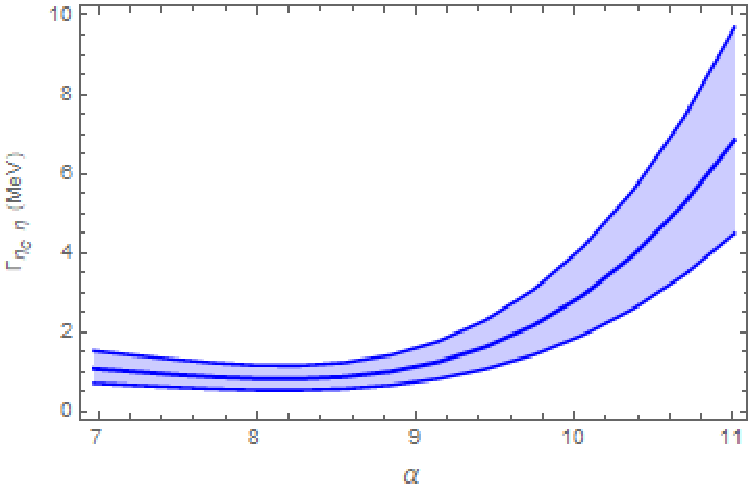}
\end{minipage}
}
\centering
\caption{The particle decay width of the $D_s\bar{D}_s$ bound state to $\eta_c\eta$ is shown for $E_b$ = 1 MeV (a), $E_b$ = 5 MeV (b), and $E_b$ = 15 MeV (c) across the allowed parameter range of $\alpha$.}
\label{EtacEta}
\end{figure}

\begin{figure}[htbp]
\centering
\subfigure[]{
\begin{minipage}[t]{0.3\linewidth}
\centering
\includegraphics[width=2.0in]{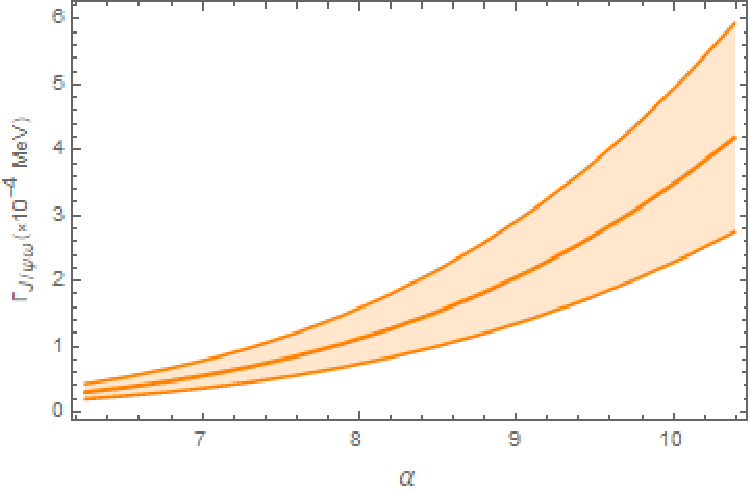}
\end{minipage}
}
\subfigure[]{
\begin{minipage}[t]{0.3\linewidth}
\centering
\includegraphics[width=2.0in]{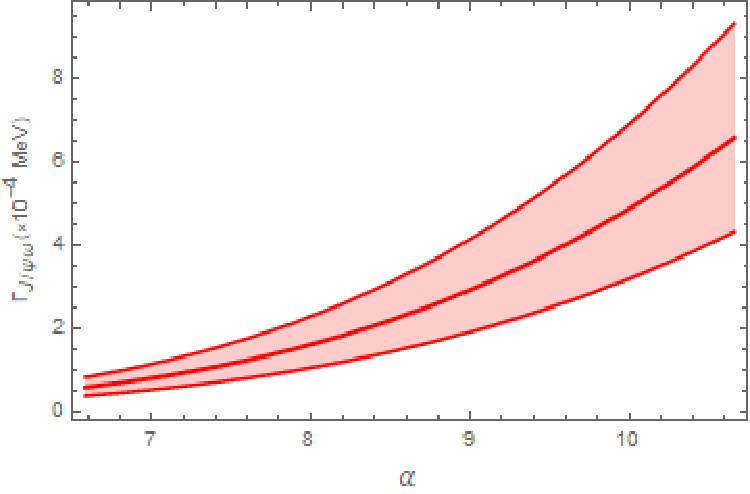}
\end{minipage}
}
\subfigure[]{
\begin{minipage}[t]{0.3\linewidth}
\centering
\includegraphics[width=2.0in]{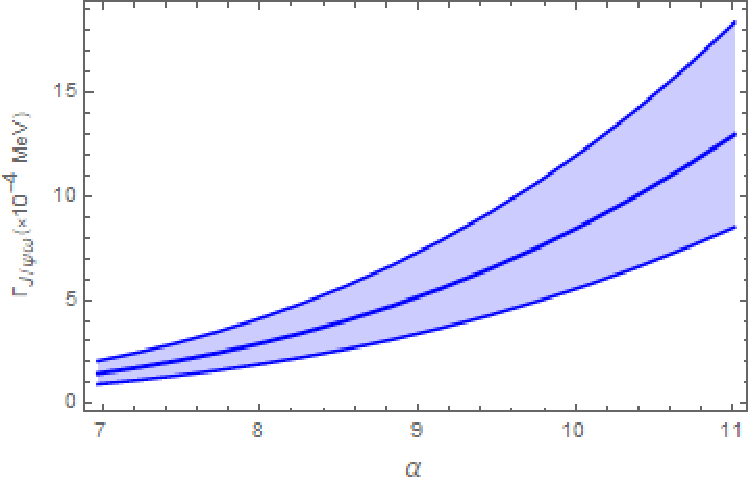}
\end{minipage}
}
\centering
\caption{The decay width of the $D_s\bar{D}_s$ bound state to $J/\psi\omega$ is shown for $E_b$ = 1 MeV (a), $E_b$ = 5 MeV (b), and $E_b$ = 15 MeV (c), across the allowed range of the parameter $\alpha$.}
\label{JpsiOmega}
\end{figure}

Based on the obtained numerical results for the $D_s\bar{D}_s$ bound state wave functions and the corresponding parameter $\alpha$, we calculate the partial decay widths of the $D_s\bar{D}_s$ bound state. The uncertainties from the coupling constants involved in the decay processes are considered. These coupling constants are derived via SU(3)-flavor symmetry, with symmetry breaking at the level of $\delta = 19\%$ \cite{Xie:2022lyw}. Following the approach in Ref. \cite{Ling:2021asz}, we estimate the uncertainty in the partial decay widths using $\Gamma = \Gamma(1+\delta)^2$. The partial decay widths for the $D_s\bar{D}_s$ bound state decaying into $D\bar{D}$, $\eta_c\eta$, and $J/\psi\omega$ final states are shown in Figs. \ref{DD}, \ref{EtacEta}, and \ref{JpsiOmega}, respectively. Our results show that these decay widths vary with the parameter $\alpha$ and are sensitive to $\alpha$ in some regions. Moreover, for a fixed value of $\alpha$, the partial decay widths of the $D_s\bar{D}_s$ bound state increase as the binding energy increases.

From Figs.~\ref{DD}, \ref{EtacEta}, and \ref{JpsiOmega}, we observe that the $D\bar{D}$ channel dominates the decays of the $D_s \bar{D}_s$ molecule, while the $J/\psi\omega$ width is the smallest. This hierarchy occurs because $D\bar{D}$ proceeds via light $K^\ast$ exchange, whose propagator is less suppressed off shell; in contrast, the $\eta_c \eta$ and $J/\psi\omega$ modes require heavy $D_s^{(\ast)}$ exchange, which introduces heavier propagators and stronger suppression. We also include $\eta$-$\eta'$ and $\omega$-$\phi$ mixing, with mixing angles $\theta_p = -19.1^\circ$ and $\theta_v = (3.4\pm0.2)^\circ$, which further suppress these two channels. The $J/\psi\omega$ channel is particularly suppressed due to the tiny $s\bar{s}$ component in $\omega$ and the small $g_{D_s^{(\ast)}D_s\omega}$ coupling constants, rendering its width orders of magnitude smaller. Additionally, for very shallow binding (1–5 MeV), we observe that $\Gamma(\eta_c \eta)$ does not increase monotonically with $\alpha$. This may be because the two Feynman diagrams ($\eta_c$ emitted from the $D_s$ versus the $\bar{D}_s$ vertex) exhibit different $\alpha$ dependences. However, it is difficult to attribute this behavior purely to kinematics or dynamics. Importantly, this nonmonotonicity occurs only for extremely shallow binding in a subleading channel and does not affect the overall decay pattern or our main conclusions. 

In Ref.~\cite{Chen:2016ncs}, the authors also investigated the decay behaviors of the $D_s^+D_s^-$ bound state using an effective Lagrangian approach. Our results for the decay widths to $D\bar{D}$ and $J/\psi\omega$ final states are consistent with those obtained in Ref.~\cite{Chen:2016ncs}. However, there is a discrepancy in the decay width to the $\eta_c\eta$ final state. In Ref.~\cite{Chen:2016ncs}, the decay width for $\eta_c\eta$ ranges from over 1 MeV to nearly 100 MeV as the parameter $\Lambda$ varies from 0.2 to 1 GeV. In contrast, in our study, the decay width for $\eta_c\eta$ varies by only a few MeV within the allowed range of the parameter $\alpha$.

Furthermore, the $X(3915)$ state, initially discovered by the Belle Collaboration~\cite{Belle:2004lle} and subsequently assigned $J^{PC} = 0^{++}$ by the BaBar~\cite{BaBar:2012nxg} and LHCb~\cite{LHCb:2020pxc} Collaborations, has been considered a possible candidate for a charmonium-like state~\cite{Guo:2012tv,Olsen:2014maa}. The mass of $X(3915)$ lies just below the $D_s\bar{D}_s$ threshold by about 16 MeV~\cite{ParticleDataGroup:2024cfk}. Ref.~\cite{Li:2015iga} suggested that the properties of $X(3915)$ can be explained if it is an $S$-wave $D_s\bar{D}_s$ bound state with a binding energy of about 18 MeV. In our model, using the average mass of $X(3915)$ from the Particle Data Group~\cite{ParticleDataGroup:2024cfk}, 3921.7 MeV, we find the partial decay widths to the $D\bar{D}$, $\eta_c\eta$, and $J/\psi\omega$ final states to be in the ranges (35.06–135.86) MeV, (0.51–3.04) MeV, and (0.49–2.05)$\times10^{-3}$ MeV, respectively. For the shallow $D_s\bar{D}_s$ bound state with $E_b = 6.2$ MeV predicted by lattice QCD, the partial decay widths to $D\bar{D}$, $\eta_c\eta$, and $J/\psi\omega$ final states are (21.92–94.77) MeV, (0.37–1.40) MeV, and (0.05–0.69)$\times10^{-3}$ MeV, respectively.

\section{Summary}

In this work, we have investigated possible bound states of the $D_s\bar{D}_s$ system using the BS equation and calculated the partial decay widths of these states to the $D\bar{D}$, $\eta_c\eta$, and $J/\psi\omega$ final states. To accomplish this, we applied the ladder and instantaneous approximations to the kernel, which includes contributions from $\phi$ and $J/\psi$ exchange diagrams. Our calculations show that the $D_s\bar{D}_s$ system can form bound states when both $\phi$ and $J/\psi$ exchange contributions are considered. However, when only one of these exchanges is included, no bound states are formed. The possible bound states of the $D_s\bar{D}_s$ system for binding energies $E_b$ = 1, 5, and 15 MeV are presented in Fig. \ref{DsDs}.

We used the numerical solutions of the BS wave functions to compute the partial decay widths of the $D_s\bar{D}_s$ bound states into $D\bar{D}$, $\eta_c\eta$, and $J/\psi\omega$ final states. The results for these partial decay widths are shown in Figs. \ref{DD}, \ref{EtacEta}, and \ref{JpsiOmega}, respectively. Our results indicate that the partial decay widths depend on the parameter $\alpha$ and are sensitive to its value in certain regions. We observe that the dominant decay channel for the $D_s\bar{D}_s$ bound state is $D\bar{D}$. Therefore, neglecting contributions from other channels such as light meson decays and photon radiation, the total decay width of the $D_s\bar{D}_s$ bound state is estimated to be in the ranges 8.56–46.39 MeV, 20.12–102.43 MeV, and 35.83–174.12 MeV for binding energies $E_b$ = 1, 5, and 15 MeV, respectively. These results show that the decay width increases with binding energy. Based on our calculations, the values of $\alpha$, and the magnitude of the decay widths, it is plausible to conclude that the $D_s\bar{D}_s$ system can form a loosely bound state. We hope that our theoretical predictions regarding the properties of the $D_s\bar{D}_s$ bound state will be tested in future experimental studies.

\acknowledgments
This work was supported by the National Natural Science Foundation of China (Project Nos. 12405115, 12275024, and 12105149 ). One of the authors (Zhen-Yang Wang) would like to thank Professors Feng-Kun Guo and Bing-Song Zou for their financial support during his visit to the Institute of Theoretical Physics, Chinese Academy of Sciences, as well as for helpful discussions with Dr. Pan-Pan Shi.

\end{document}